\documentclass[journal]{IEEEtran}

\usepackage{xcolor}
\usepackage{cite}
\usepackage{soul}
\usepackage{tikz}
\usetikzlibrary{matrix,calc,shapes,arrows,positioning}
\pgfdeclarelayer{background}
\pgfdeclarelayer{foreground}
\pgfsetlayers{background,main,foreground}

\usepackage[cmex10]{amsmath}
\usepackage{amsmath,amsthm,mathtools}
\usepackage{amssymb}
\ifCLASSOPTIONcompsoc
  \usepackage[caption=false,font=normalsize,labelfont=sf,textfont=sf]{subfig}
\else
  \usepackage[caption=false,font=footnotesize]{subfig}
\fi

\usepackage{siunitx}
\usepackage{ifthen}
\usepackage{calc}
\usepackage{makeidx}
\usepackage{nomencl}
\usepackage{multirow}
\usepackage{footnote}
\usepackage{euscript}

\makenomenclature
\nomenclature[aa]{$\mathcal{B}$}{Nomenclature to be filled in later}
\makeatletter
\def\BState{\State\hskip-\ALG@thistlm}
\makeatother
\begin{document}

\title{A Nonparametric Bayesian Methodology for Synthesizing Residential Solar Generation and Demand Data}

\author{Thomas~Power,~\IEEEmembership{Student Member,~IEEE,}
Gregor~Verbi\v{c},~\IEEEmembership{Senior Member,~IEEE,}
Archie~C.~Chapman,~\IEEEmembership{Member,~IEEE,}	
\thanks{Thomas~Power, Gregor~Verbi\v{c}, and Archie~C.~Chapman are with the School of Electrical and Information Engineering, The University of Sydney, Sydney, New South Wales, Australia. e-mails: {tpow5976, gregor.verbic, archie.chapman}@sydney.edu.au.}
\thanks{}}

\maketitle

\begin{abstract}
The uptake of \emph{behind-the-meter} distributed energy resources in low-voltage distribution networks has reached a level where network issues have started to emerge, which requires new tools for operation and planning. 
In this paper, we propose a methodology for synthesizing stochastic demand and generation profiles for unobserved customers with rooftop PV, called \textit{prosumers}.
The proposed model bridges the gap between the limited available empirical data, and the large amount of high-quality, stochastic demand and generation data required for probabilistic analysis.
The approach employs clustering analysis and a Dirichlet-categorical hierarchical model of the features of unobserved prosumers. 
Based on the data of clusters of prosumers, Markov chain models of demand and generation profiles are constructed from empirical data, and synthetic demand profiles are subsequently sampled from these. 
The sampled traces are cross-validated and show a good statistical fit to the observed data. Two case studies are considered to confirm the validity of the proposed methodology. The first studies the impact of behavioral differences on the synthetic demand profiles, while the second looks at the impact of varying solar generation penetration on demand profiles.
\end{abstract}


\IEEEpeerreviewmaketitle

\section{Introduction}
\IEEEPARstart{T}{HE} emergence of cost-effective behind-the-meter distributed energy resources, including on-site generation, energy storage, electric vehicles, and flexible loads is changing the way that electricity consumers source and consume electric power.
To give some context, installed capacity of rooftop PV has increased globally from approximately \SI{4}{\giga \watt} in 2003 to nearly \SI{227}{\giga \watt} in 2015, driven by government incentives, electricity price increases, and decreasing PV capital cost \cite{IEA_2018}.
In Australia, the Energy Networks Australia and the Australian Commonwealth Scientific and Industrial Research Organisation (CSIRO) have estimated the projected uptake of solar PV and battery storage in 2050 to be \SI{80}{\giga \watt} and \SI{100}{GWh} \cite{CSIRO_ENA_2017}, respectively, which will represent between 30\%--50\% of total demand, a scenario called ``Rise of the Prosumer''. They refer to the \emph{prosumer} as a small-scale (residential, commercial and small industrial) electricity consumer with on-site generation.

The rising penetration of PV and battery storage, however, presents a number of challenges with regard to the operation of low-voltage networks. Most existing studies have only focused on the impact of rooftop PV on voltages. Chen et al. \cite{R12}, for example, investigated the impact of an increasing penetration of distributed generation on the voltage profiles of US distribution networks, while Widen et at. did that for a Swedish network \cite{R14}. A recent UK study by Navarro-Espinosa and Ochoa in the UK \cite{R13} considered a wider range of \emph{low carbon technologies}, including battery storage. They didn't however, consider battery scheduling as a result of \emph{home energy management} \cite{GVHT, adplearn}.

A common feature of these studies is that they use probabilistic analysis to properly account for the variation in both PV generation and electricity demand. The latter, in particular, is extremely important in the context of home energy management, as it can have a significant impact on the battery schedule and, consequently, the grid demand. Against this backdrop, this paper proposes a methodology for synthesizing statistically representative demand and PV generation profiles using only a limited amount of observed smart meter data. 

\subsection{Related work on Synthetic Demand and Generation Profiles}
Much of the previous study on the synthesis of stochastic demand and generation profiles has centered around a 'bottom-up' approach. This particular framework is utilized by Widen, Nilsson and Wackelgard \cite{R27} to model building occupancy for the purposes of forecasting lighting demand. Their methodology involves simulating building occupancy as a Markov chain, defined by a state transition probability matrix. Richardson, Thompson and Infield \cite{R30} and Page et al. \cite{R31} use a similar methodology for sampling building occupancy profiles.
McKenna and Thompson extend this methodology \cite{R32} in their high-resolution stochastic integrated thermal-electrical domestic demand model (the \textit{CREST} model).
Their particular occupancy model is central to the determination of a residence’s demand profile, and allows for a number of distinct day types, whereby it defines separate transition probability matrices for both weekdays and weekends, which captures weekly behavioral patterns. This is effective in that there are significant structural differences in the behavior of household demand between weekends and weekdays.
The generation component in the CREST model first involves sampling the weather conditions. Similarly to the demand model, a transition matrix is constructed from prior historical data for the clearness index, which is a measure of the extent to which incident irradiance is blocked from reaching a PV module by cloud cover. In a method similar to that used by Hofmann et al. \cite{R35} and Bright et al. \cite{R36}, initial states are chosen at random, and the clearness index profile is sampled at each timestep through a repeated application of a state transition probability matrix \cite{R32}. This is then used to generate stochastic PV generation profiles \cite{R32}.

\subsection{Research Overview}
This work develops a framework for producing a large collection of residential demand and generation profiles from limited observed \emph{smart meter} data. Previous studies have either involved using a simplified probabilistic model \cite{R12}, or reconstructing profiles using highly granular observed data, which is computationally costly \cite{R32}. The methodology described in this paper requires data on a household level that is somewhat less granular, and more readily accessible. It also aims to retain the variance that is normally observed among observed demand and generation profiles. 

The rest of the paper proceeds as follows. Section II details the formulation of the proposed model, while Section III cross-validates the model. Section IV outlines potential applications of the model through two specific case studies. Finally, Section V concludes.

\section{Model Formulation}
We wish to synthesize typical solar generation and demand profiles for a household according to specific features.
These features can either be continuous or discrete, and may not be independently distributed within the data. 
Discrete features generally need no treatment prior to assignment. 
However, if the number of values taken on by numerical features is impractically large, some clustering analysis may be done, as described next.
After this, the method used to identify a Markov chain model of the demand and solar generation profiles is described, followed by the steps used to generate new synthetic demand and generation traces.

\subsection{Feature-based Cluster Assignment}
For this clustering analysis, empirical data was collected during the \textit{Ausgrid Smart-Grid Smart-City} (SGSC) project \cite{R46}.
 Clustering is important because (i) considering each customer as a single category is computationally expensive, and (ii) it provides generalizable statistical information as the demand and PV generation in each set are correlated with their features. 

Let $n \in \mathcal{N}$ 
and $m \in \mathcal{M}$ denote the set of observed and unobserved customers, respectively. 
Clustering analysis is run on empirical data to assign the $n \in \mathcal{N}$ customers into representative sets, denoted $k \in \mathcal{K}$ according to their features.
The features of demand are the day types (weekday or weekend) and number of residents, while those for PV include the PV capacity, panel orientation and weather information.
Clustering is completed using either $k$-means clustering \cite{mac_k} or \textit{maximum a-posteriori Dirichlet process mixtures} (MAP-DP) clustering \cite{R25}, which is useful for instances in which the number of clusters cannot be easily determined.  

\subsection{Estimating the Dirichlet Distribution}
After clustering, we could compute the frequencies, 
$\{p_s\}_{k \in \mathcal{K}}$, of each observed customer being a member of a certain cluster. 
These values can be interpreted as the probability of a new, unobserved customer having certain features. However, they are only an estimate across the observed customers, and directly using them to allocate features fails to properly consider the error in this estimate, which can be significant where the fraction of customers observed is small. Thus, a Bayesian estimation approach is employed.

Specifically, the model uses the count of each $k \in \mathcal{K}$ in the observed $\mathcal{N}$ as a hyper-parameter of a  Dirichlet distribution, which itself is sampled to yield a  \textit{categorical} probability distribution over the features for unobserved customers, ${m \in \mathcal{M}}$.  
Formally, this is given by: 
\[ {\displaystyle {\begin{array}{rcl}
\boldsymbol{\alpha }& & \text{Vector of cluster counts}\\
\mathbf{q} \mid \boldsymbol{\alpha} & \sim &\operatorname{Dir} (\boldsymbol {\alpha })\\
S_m \mid \mathbf {q} & \sim &\operatorname{Cat}(\mathbf{q} )
\end{array}}}
\]

In more detail, given the clusters, we use the Dirichlet distribution, given by the the probability density function:
\begin{align}
f(\textbf{q}_j;\boldsymbol{\alpha}_j)=\frac{\Gamma (\alpha_{j0})}{\prod_{k=1}^{m}\Gamma(\alpha_{jk})}\prod_{k=1}^{m}q_{kj}^{\alpha_{jk} -1}.
\end{align}
The Dirichlet distribution accounts for variance in the distribution of features across unobservable customers, which overcomes a limitation of previous models.

In this case, $\textbf{q}_j$ is the resultant probability measure for feature $j$ drawn from the Dirichlet distribution, with individual elements $q_{jk}$, where $k \in \{1,2,...,m\}$, and $\sum_{k=1}^{m}q_k = 1$, while the corresponding random probability measures and elements are given by $\textbf{Q}_j$ and $Q_{jk}$, respectively. The distribution is parametrized by $\boldsymbol{\alpha}_j$ which is given by:
\begin{align}
\alpha_{jk} = \sum_{i = 1}^{n}\delta_{c^{'}_{ij},k}, \quad \forall j \leq u, \quad \forall k \leq m,
\end{align}
where $\alpha_{j0} = \sum_{k=1}^{m}\alpha{jk}$, and $\delta$ is the Kronecker delta.
For each feature, $\textbf{q}_j$ parametrises a draw from a categorical distribution, such that for $n^{\ast}$, unobserved prosumers, we have:
\begin{align}
	\textbf{A}^{\ast}_{j} = [A^{\ast}_{j1}, A^{\ast}_{j2}, ..., A^{\ast}_{jm}],
\end{align}
which is defined by the probability density function:
\begin{align}
f(\textbf{A}_j^{\ast};\textbf{q}_j) = \frac{\Gamma \big{(}\sum_{k=1}^{m}A_{j}+1 \big{)}}{\prod_{k=1}^{m}\Gamma(A_{jk}+1)}\prod_{k=1}^{m}\textbf{p}_{k}^{A_{jk}}.
\end{align}
This collection of feature counts defined by $\textbf{A}_{j}$ are then assigned to the unobserved prosumers.

\subsection{Demand Model}
The demand model outlined in Fig. \ref{fig:flowchart} synthesizes stochastic demand profiles for unobserved prosumers using observed data for characteristically similar prosumers. To do this, Markov chain modeling is used, with a state transition matrix defined for each timestep throughout the day. This model is time inhomogeneous, as there are structural differences in the behavior of household energy usage at different points in the day. Once the timestep length and the collection of relevant observable customers have been defined, the state transition matrices are then constructed. For this study, a timestep of 30 minutes is used and as such, 48 different state transition matrices will be required to sample a profile over a 24 hour period. In addition to this, it will need to be specified whether the resulting profile is to be synthesized for a weekend or a weekday, as they have significantly different behavior. Following this, the state transition matrices can be defined:
\begin{align}
\textbf{M} &= [\textbf{M}_1, \textbf{M}_2,...,\textbf{M}_{48}].
\end{align}
In each matrix $\textbf{M}_k$, the states are defined for
\begin{align}
m_{i,j,k}, \quad i,j \in \{0,1,...,n\},
\end{align}
and are used to represent the amount of energy consumed for the duration of timestep $k$. In the case of the validation data used in this work \cite{R46}, the data is logged in terms of the number of \SI{}{kWh} consumed in each relevant 30 minute period. In order to construct the state transition probability matrices, these readings are discretized with each \SI{}{kWh} reading rounded to the nearest \SI{0.01}{kWh}, and the possible attainable states are defined as being this value multiplied by 100. The state, therefore, of a reading of \SI{1.023}{kWh} will be 102. These states are defined up to an arbitrary maximum limit $n$, which is set well above the maximum observed state.
The state transition probability matrices are built up from the observed data by taking:
\begin{align}
m_{i,j,k} = \sum_{h = 1}^{N}\delta_{x_{hk-1},i} \delta_{x_{hk},j},
\end{align}
over each similar observed prosumer in the available data. Stochastic demand profiles can then be directly sampled from this set of matrices by first calculating the conditional distribution of the initial state, which is given by:
\begin{align}
\textbf{p} = [p_1, p_2, ..., p_n], \quad p_j = \frac{\sum_{i = 1}^{n}m_{i,j,1}}{\sum_{i = 1}^{n}\sum_{j = 1}^{n}m_{i,j,1}}.
\end{align}
Once this initial state $s_1$ is drawn from the categorical distribution parametrized by $\textbf{p}$, subsequent states for the remainder of the day can be drawn from the following matrices. For each state $s_k$, the associated probability measure is derived from $\textbf{M}_{s_{k-1},j,k}$ using kernel density estimation. Gaussian kernel density estimation is used to make each state attainable, even if it has not been observed in the data. It is given by:
\begin{align}
p_{s_{k-1},j,k} = \frac{\sum_{j^{\ast}=1}^{n}m_{s_{k-1},j^{\ast},k} e^{-\frac{(j-j^{\ast})^{2}}{2h^2}}}{\sum_{j^{\ast} = 1}^{n} e^{-\frac{(j-j^{\ast})^{2}}{2h^2}}}.
\end{align}
From this kernel density estimate, the state $s_k$ can be drawn from a categorical distribution:
\begin{align}
s_k \sim \text{Cat}(\textbf{p}_{s_{k-1},k}).
\end{align}
Before synthesizing these profiles, it will need to be specified whether the the profiles are to be synthesized for a weekday or a weekend. There are significant structural differences in energy usage behavior between weekdays and weekends \cite{IHMM7}, and as such, only observed data from weekdays is to be used to synthesize weekday profiles, while only observed data from weekends is to be used to calculate weekend profiles.

\tikzstyle{decision} = [diamond,aspect=3, draw, 
    text width=15em, fill=blue!20, text badly centered, node distance=3cm, inner sep=0pt]

\tikzstyle{block} = [rectangle, draw, minimum width=0.4*\columnwidth,
     fill=blue!20, text centered, rounded corners, minimum height=2em,text width=0.8*\columnwidth]

\tikzstyle{line} = [draw, -latex', thick]

\tikzstyle{cloud} = [draw,fill=red!20, ellipse, node distance=3cm,
    minimum height=2em]

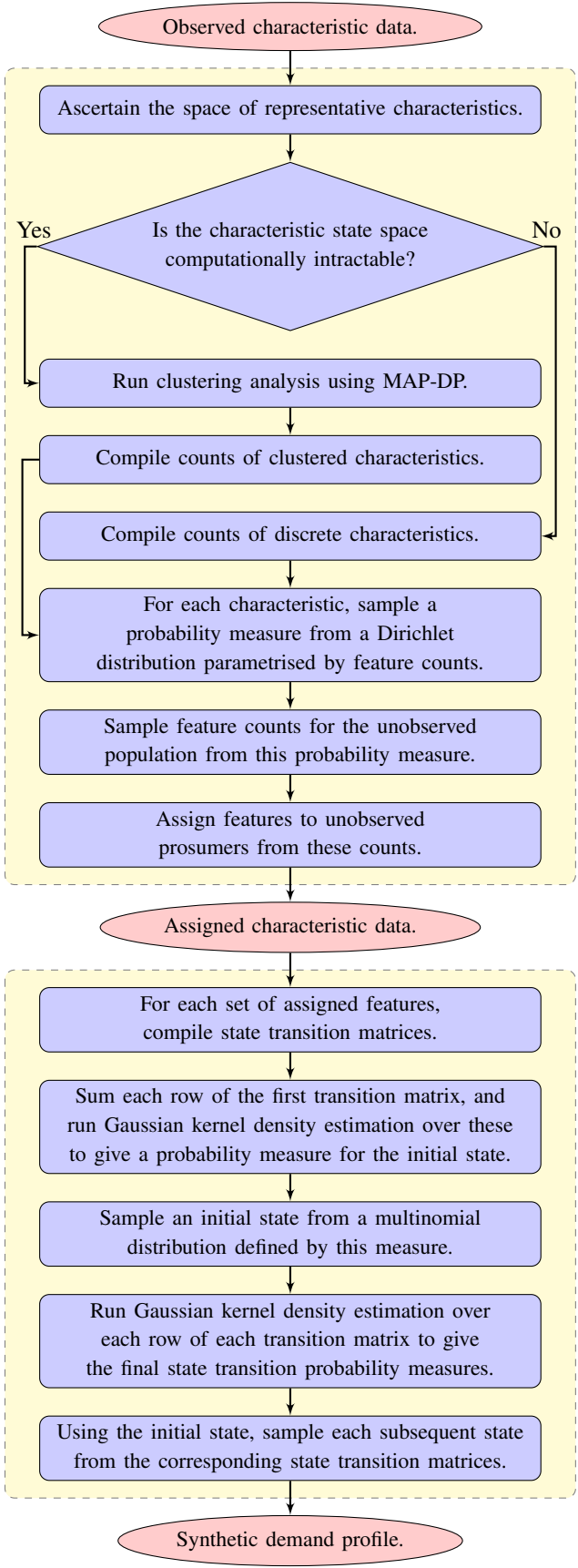
\begin{figure}
\centering
\begin{tikzpicture}[node distance = 2cm,auto]
	\node [cloud](init){\small{Observed characteristic data.}};
	
    \node [block, below = 0.5cm of init](charac){\small{Ascertain the space of representative characteristics.}};
	
    \node [decision, below = 0.4cm of charac](intractable){\small{Is the characteristic state space computationally intractable?}};
    
    \node [block, below = 0.4cm of intractable](map-dp){\small{Run clustering analysis using MAP-DP.}};

	\node [block, below = 0.4cm of map-dp](clustered){\small{Compile counts of clustered characteristics.}};
    
    \node [block, below = 0.4cm of clustered](discrete){\small{Compile counts of discrete characteristics.}};
     
    \node [block, below = 0.4cm of discrete](dirichlet){\small{For each characteristic, sample a probability measure from a Dirichlet distribution parametrised by feature counts.
}};

	\node [block, below = 0.4cm of dirichlet](feature){\small{Sample feature counts for the unobserved population from this probability measure.}};
    
    \node [block, below = 0.4cm of feature](assign){\small{Assign features to unobserved prosumers from these counts.}};
    
    \node [cloud, below = 0.5cm of assign](unobserved){\small{Assigned characteristic data.}};

	\node [block, below = 0.5cm of unobserved](compile){\small{For each set of assigned features, compile state transition matrices.}};

	\node [block, below = 0.4cm of compile](initial){\small{Sum each row of the first transition matrix, and run Gaussian kernel density estimation over these to give a probability measure for the initial state.}};
    
    \node [block, below = 0.4cm of initial](state){\small{Sample an initial state from a multinomial distribution defined by this measure.}};
    
        \node [block, below = 0.4cm of state](kde){\small{Run Gaussian kernel density estimation over each row of each transition matrix to give the final state transition probability measures.}};
    
    \node [block, below = 0.4cm of kde](sample){\small{Using the initial state, sample each subsequent state from the corresponding state transition matrices.}};
    
    \node [cloud, below = 0.5cm of sample](profile){\small{Synthetic demand profile.}};
    
	\path [line] (init) -- (charac);
    \path [line] (charac) -- (intractable);
    \path [line] (intractable.west)--node [above, near start] {Yes} ++(-5pt,0pt)  |- (map-dp.west);
    \path [line] (intractable.east)--node [above, near start] {No} ++(5pt,0pt) |- (discrete.east);
    \path [line] (map-dp) -- (clustered);
    \path [line] (clustered.west)--++(-7pt,0pt) |-(dirichlet.west);
    \path [line] (discrete)--(dirichlet);
    \path [line] (dirichlet)--(feature);
    \path [line] (feature)--(assign);a
    \path [line] (assign)--(unobserved);
    \path [line] (unobserved)--(compile);
    \path [line] (compile)--(initial);
    \path [line] (initial)--(state);
    \path [line] (state)--(kde);
    \path [line] (kde)--(sample);
    \path [line] (sample)--(profile);
    
\newcommand{\background}[5]{%
  \begin{pgfonlayer}{background}
    \path (#1.west |- #2.north)+(-0.5,0.25) node (a1) {};
    \path (#3.east |- #4.south)+(+0.5,-0.25) node (a2) {};
    \path[fill=yellow!20,rounded corners, draw=black!50, dashed]
      (a1) rectangle (a2);
      \path (#3.east |- #2.north)+(0,0.25)--(#1.west |- #2.north) node[midway] (#5-n) {};
      \path (#3.east |- #2.south)+(0,-0.35)--(#1.west |- #2.south) node[midway] (#5-s) {};
      \path (#3.east |- #2.north)+(0.7,0)--(#3.east |- #4.south) node[midway] (#5-w) {};
  \end{pgfonlayer}}
    
    \background{map-dp}{charac}{discrete}{assign}{bk1}
    \background{initial}{compile}{kde}{sample}{bk2}
       
\end{tikzpicture}
\caption{The characteristic assignment and profile synthesis process.} \label{fig:flowchart}
\end{figure}

\subsection{Generation Model}
The generation model is the third key component in this model. It is intended to synthesize solar generation data in order to compile stochastic daily profiles similar to the demand model mentioned previously. This generation model was developed and described in detail in a previous work \cite{GV_TP_AUPEC}, and utilizes a Markov chain framework similar to that provided by the CREST model \cite{R30}.
In this model, solar PV generation is modelled using the following formula:
\begin{align}
P = \eta \times A \times TIF \times CI \times G,
\end{align}
where $\eta$ is the efficiency of the PV module, and $A$ is the area of the module. For the purposes of this work, $TIF$ is the \lq time irradiance factor\rq , and is a measure of the incident irradiance on a PV module with respect to the latitude of the panel, orientation, time of day and time of year. $G$ is the extraterrestrial irradiance constant, while $CI$ is the clearness index, which represents the extent to which cloud cover obstructs irradiance of the panel. This is the stochastic quantity that this model emulates. As detailed in previous work \cite{GV_TP_AUPEC}, the clearness index for a particular prosumer is sampled from a single state transition probability matrix, which is constructed from observed generation data. This is substituted into (20) to obtain stochastic PV generation values for each timestep of the day.

\section{Model Validation}
\subsection{Feature Assignment}
The following section details the results of a cross validation of the model using data sourced from Ausgrid's  Smart Grid, Smart City (SGSC) data set\cite{R46}. This data contains a three year sample of half hourly household demand and generation measurements for approximately 13,000 households. Following model cross validation, a number of case studies are then conducted in Section IV to demonstrate potential applications. 
\subsection{Cross Validation}
\begin{figure}
\vspace{0.0cm}
\includegraphics[width= \columnwidth]{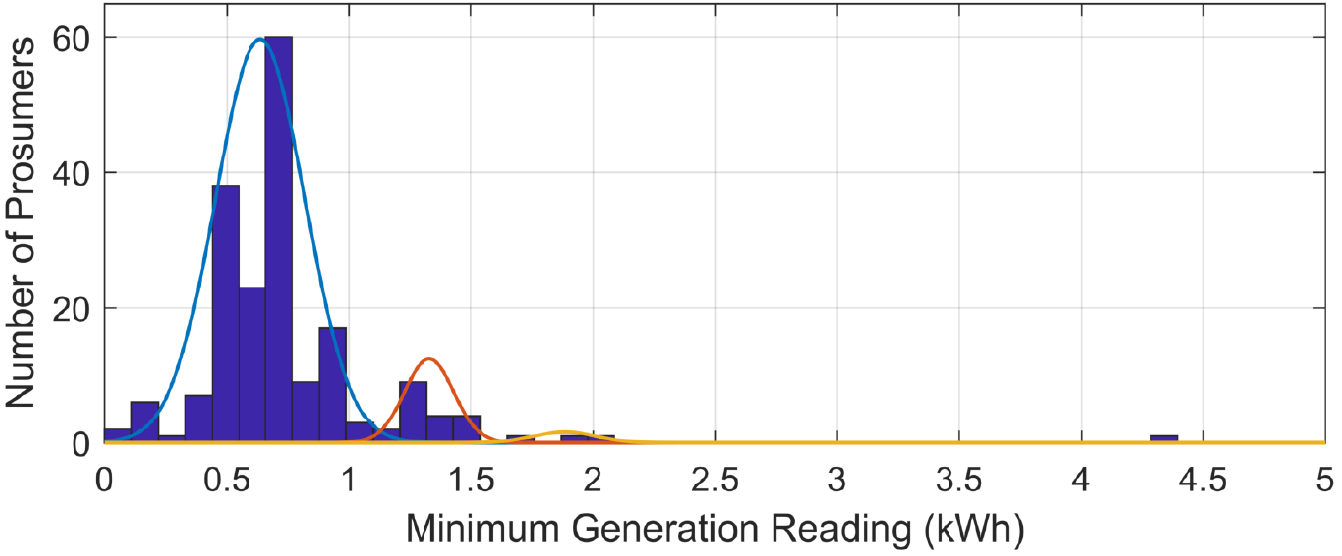}
\caption{Histogram of peak solar readings for all observed generators.} \label{fig:caps}
\vspace{0.0cm}
\end{figure}

The first part of this framework involves the clustering and assignment module. For the SGSC data set, the peak readings for solar installations are the only relevant variable available for clustering, as they act as a proxy for the capacity of the system. The distribution of the 189 observations used for clustering is shown in Fig. \ref{fig:caps}.

One of the limitations of $k$-means clustering is that the number of clusters and prior estimates for cluster locations are required before running the algorithm. 
Therefore we use the MAP-DP clustering algorithm proposed by Raykov et al. \cite{R25}. For this algorithm, an initial cluster mean and variance is specified, along with a prior estimate for the variance of new clusters and the Dirichlet parameter $\alpha$, which represents the propensity of the algorithm to pick new clusters for assignments. In the case of the peak solar reading data, this clustering algorithm was run with the naive prior cluster mean of 0.7428, and variance of 0.1700.

Using these values, the MAP-DP algorithm was run using a Dirichlet parameter of 9, which yielded the 4 clusters shown in Table \ref{tab:mapdp9}. The error is comparable to k-means clustering, which used an informed prior. This therefore demonstrates the efficacy of MAP-DP under a lack of prior information.

\begin{table}
\begin{center}
\caption{Results of MAP-DP clustering with a Dirichlet parameter of 9} \label{tab:mapdp9}
\begin{tabular}{|c|cccc|c|}
\hline
\textbf{Cluster ($i$)} & 1 & 2 & 3 & 4 & \textbf{Error}\\
\hline
\textbf{Location ($\bar{x}_i$)}&0.6368&1.3293&1.8843&4.3960&-\\
\textbf{Variance ($s_i$)}&0.0338&0.0104&0.016&0&- \\
\textbf{Population ($n_i$)}&165&20&3&1&5.83\\
\hline
\end{tabular}
\end{center}
\end{table}
A superposition of these clusters comprises a Gaussian mixture model, which is then normalized for assignments to unobserved prosumers, as shown in Fig.~2. The Gaussian mixture model is given by the following probability mass function:
\begin{align}
f(x) = \frac{\sum_{i=1}^{N}\frac{1}{\sqrt{2 \pi s_{i}^{2}}}e^{-\frac{(x-\bar{x}_i)^2}{2s_{i}^{2}}}}{\sum_{i=1}^{N}n_i},
\end{align}
where $\bar{x}_i$ is the mean, $s_i$ is the variance and $n_i$ is the population of cluster i. The MAP-DP clustering analysis has therefore revealed a set of three Gaussian distributed clusters from which this peak solar data is distributed. In contrast to the $k$-means algorithm, this analysis did not require prior knowledge of cluster centroids or variances, but only an estimate for the Dirichlet parameter. This makes the MAP-DP algorithm much more versatile where little is known about the subject data.
Once these features are clustered, the resulting set of clusters can be used to parametrize a Dirichlet distribution. This defines a distribution for the probability measure for the assignments, thereby giving it a variance as well as an expected value, as illustrated in Table \ref{tab:dirdraw}.

\begin{table}
\begin{center}
\caption{A sequence of draws from a Dirichlet distribution parametrised by MAP-DP clustered peak solar output data} \label{tab:dirdraw}
\begin{tabular}{|c|cccc|}
\hline
\textbf{Draw} & $q_1$ & $q_2$ & $q_3$ & $q_4$ \\
\hline
1 & 0.8802 & 0.1104 & 0.0049 & 0.0046 \\
2 & 0.8370 & 0.1503 & 0.0106 & 0.0021 \\
3 & 0.8567 & 0.1237 & 0.0053 & 0.0143 \\
4 & 0.8990 & 0.0925 & 0.0079 & 0.0005 \\
5 & 0.8426 & 0.1459 & 0.0088 & 0.0027 \\
6 & 0.8691 & 0.1006 & 0.0197 & 0.0105 \\
7 & 0.8819 & 0.0805 & 0.0321 & 0.0054 \\
8 & 0.8758 & 0.0967 & 0.0239 & 0.0036 \\
9 & 0.8042 & 0.1659 & 0.0211 & 0.0087 \\
10 & 0.8539 & 0.1270 & 0.0137 & 0.0054 \\
\hline
\textbf{Mean} & 0.86004 & 0.11935 & 0.01480 & 0.00578 \\
\hline
\textbf{Variance} & 0.00075 & 0.00079 & 0.00008 & 0.00001 \\

\hline
\end{tabular}
\end{center}
\end{table}
Following the sampling from a Dirichlet distribution, the feature counts drawn from a categorical distribution parametrized by these samples reflect the mean and variance of the probability measure itself, as shown in Table \ref{tab:dirdraw}, which emulates the variance in the estimation of the distribution of features over the whole population. 

\subsection{Demand Model Validation}
\begin{figure}
\centering
\vspace{0.0cm}
\includegraphics[width= \columnwidth]{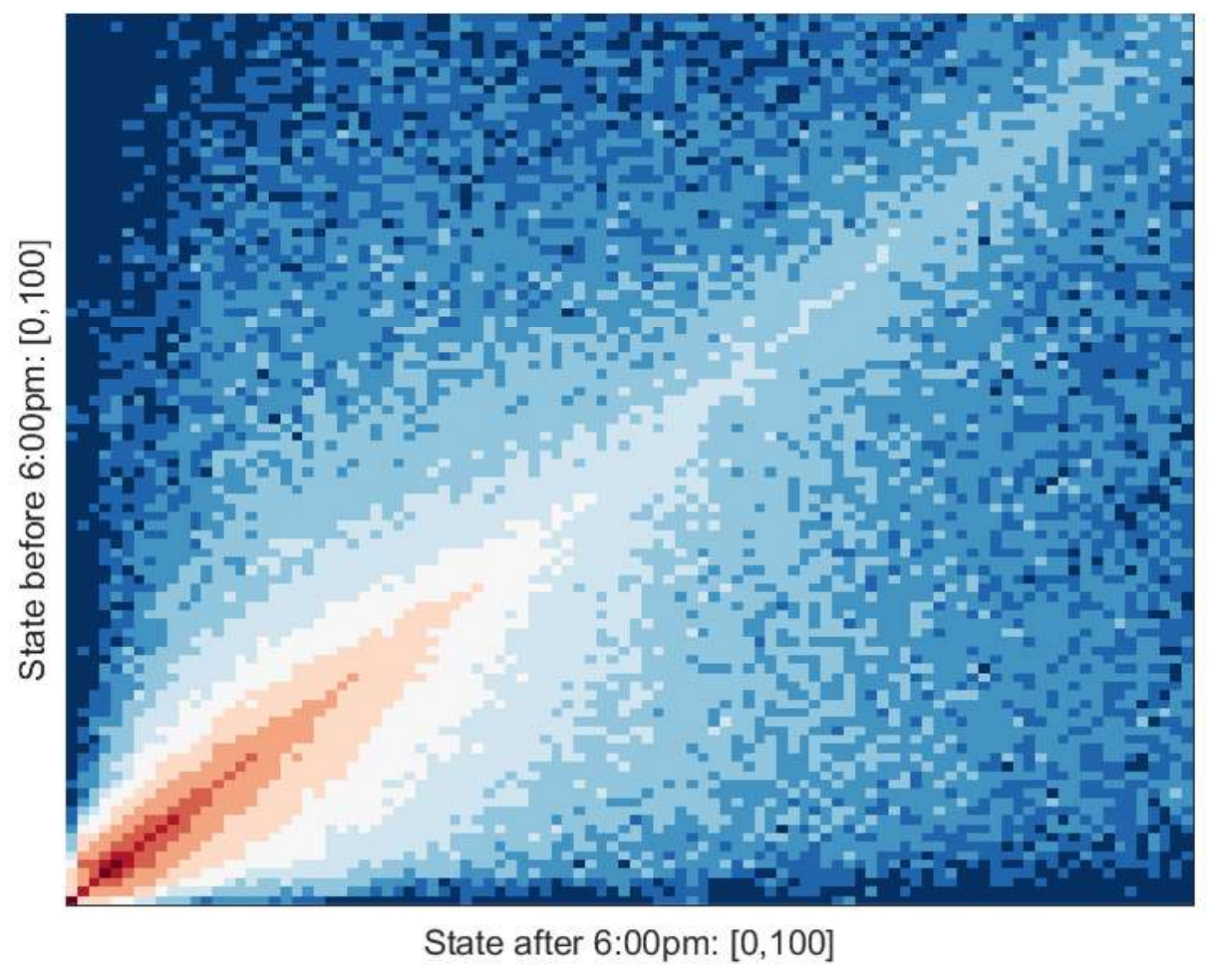}
\caption{Logarithmic heatmap of the Markov state transition matrix calculated at 6:00pm for 1000 prosumers.} \label{fig:HM_c}
\vspace{0.0cm}
\end{figure}

The function of the demand module in this model is to take a collection of demand data from observed prosumers as an input, and to synthesize either daily or multiple day profiles based on this data. Following the assignment of features to unobserved prosumers, the set $\textbf{M}$, of state transition matrices can then be populated using observed data from prosumers sharing similar features. In this case, the set of matrices is limited to $700 \times 700 \times 48$, which corresponds to the maximum demand being \SI{7}{kWh} over a half-hour period. In order to conduct a cross-validation of this model, the synthetic profiles are compared to those observed from a separate subset of the data to check for similarity. In order for this cross validation to be valid, subsets of the data need to be taken such that they do not contain any nested statistical factors. As such, the data random selection of 1000 prosumers are used for the model construction, and a separate random subset of another 1000 prosumers is used for cross validation. The matrices calculated using this process are large and sparse, and as such, and as such, the state transition matrix at 6:00pm on weekdays is visualized for states 0 to 100 as a heatmap in Fig. \ref{fig:HM_c}.

As these matrices are sparse, the following logarithmic transform of the counts is used for $m_{ijk} \in M_i$:
\begin{align}
m^{*}_{ijk} = \text{log}_{10}|m_{ijk}| \quad i,j \in \{1,2,...,100\}.
\end{align}

The heatmap shown in Fig. \ref{fig:HM_c}, uses red to denote a high count, while blue indicates a low count. Each pixel represents a distinct state transition count, with the previous state, or state before 6:00pm denoted by the row, and the state after 6:00pm denoted by the column.

\begin{figure}
\centering
\includegraphics[width= \columnwidth]{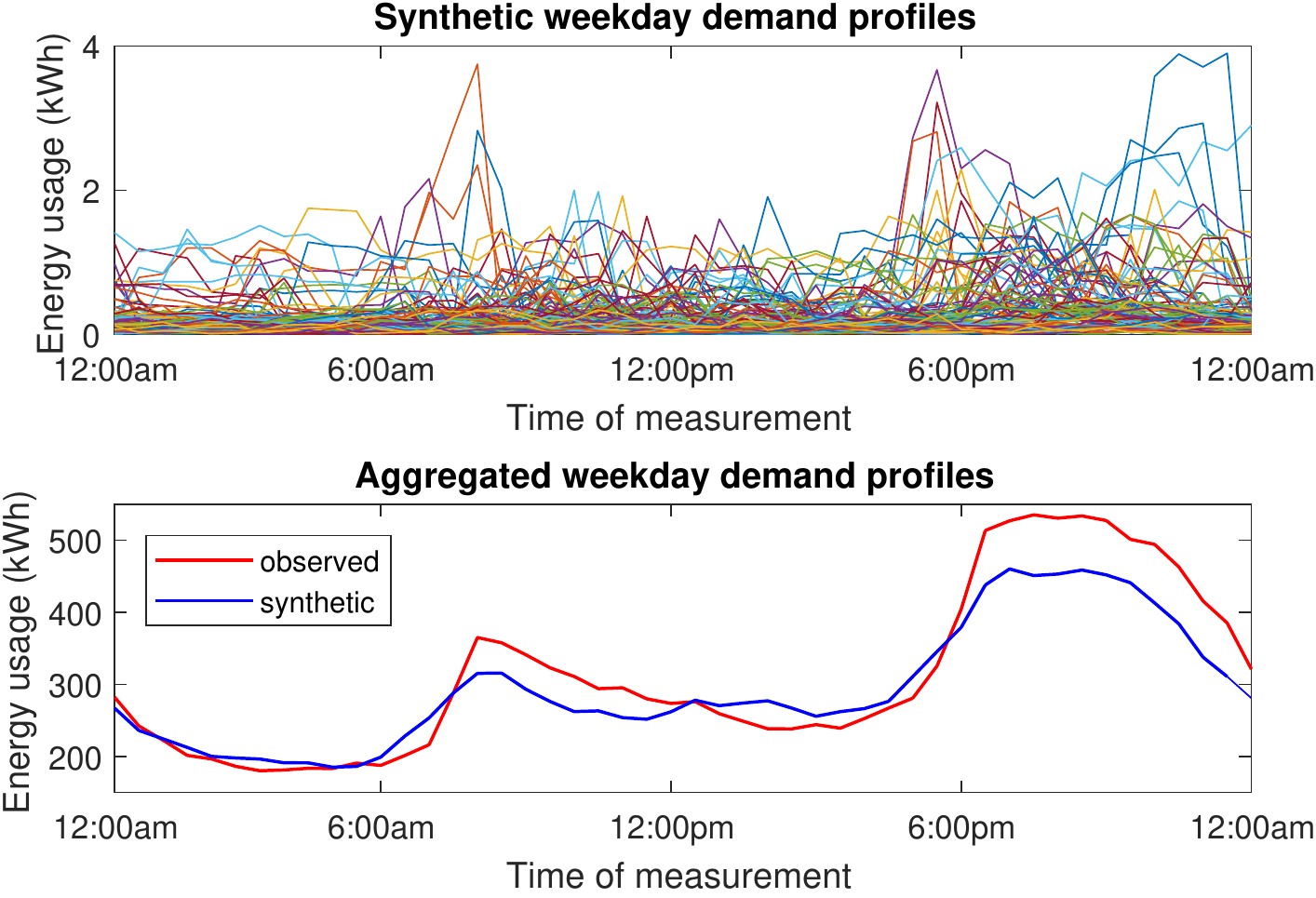}
\caption{Sample synthetic demand profiles (top) and their average value compared to the observed average profile (bottom).} \label{fig:profiles}
\end{figure}

It can therefore be seen from Fig. \ref{fig:HM_c} that a majority of transitions take place from states near zero, to similar states near zero. There are two major features to this data, the first of which is that transitions to higher states tend to be less likely than transitions to lower states. The second feature of this visualization is that transitions to similar states are more likely than transitions to dissimilar states. From this set of matrices, profiles can be sampled directly, as shown in Fig. \ref{fig:profiles} (top). When aggregated, however, it is apparent that the mean of this synthetic data matches that of the observed data. This is illustrated in Fig. \ref{fig:profiles} (bottom). 
It is visually clear that the data synthesized by the model shows similar behavior to the observed data.

In order to better quantify this, the mean absolute error of the estimate given by the synthetic profiles is calculated. This is given by:
\begin{align}
\text{Mean Absolute Error} = \frac{1}{48}\sum_{i=1}^{48}\frac{|x_i-x_{i}^{*}|}{x_i},
\end{align}
where 48 is the number of samples in the daily profile, $x_i$ is the observed value at timestep $i$, and $x^{*}_{i}$ is the synthetic value at timestep $i$. The mean absolute error is calculated for the weekday profile to be 9.80\%. Given that the samples range between approximately \SI{200}{kWh} and \SI{500}{kWh} over the course of a day, this error rate is small compared to the range of the observed data itself. It is therefore evident that the model can accurately represent the behavior of observed data. It is also evident that a large amount of information is lost when modeling is done using the aggregate, as the information on the variability of individual profiles is not retained.

\subsection{Multiple Day Profiles}
One important aspect of the Dirichlet process is its reinforcing property as detailed by MacQueen \cite{mac_k}. This allows for the generation of samples that have a tendency to follow recurring trajectories and sequences of states. In this way, states and sequences of states that have previously been sampled more often have a higher chance of being sampled again, and recurring behaviors will start to arise. To emulate the recurring behavior of an actual consumer, we propose a variation of the process described in Section II-C. This process transforms the original matrices of counts uniquely for each prosumer to be modeled, by drawing from a Dirichlet distribution parametrized by each row.

\begin{figure}
\centering
\vspace{0.1cm}
\includegraphics[width=\columnwidth]{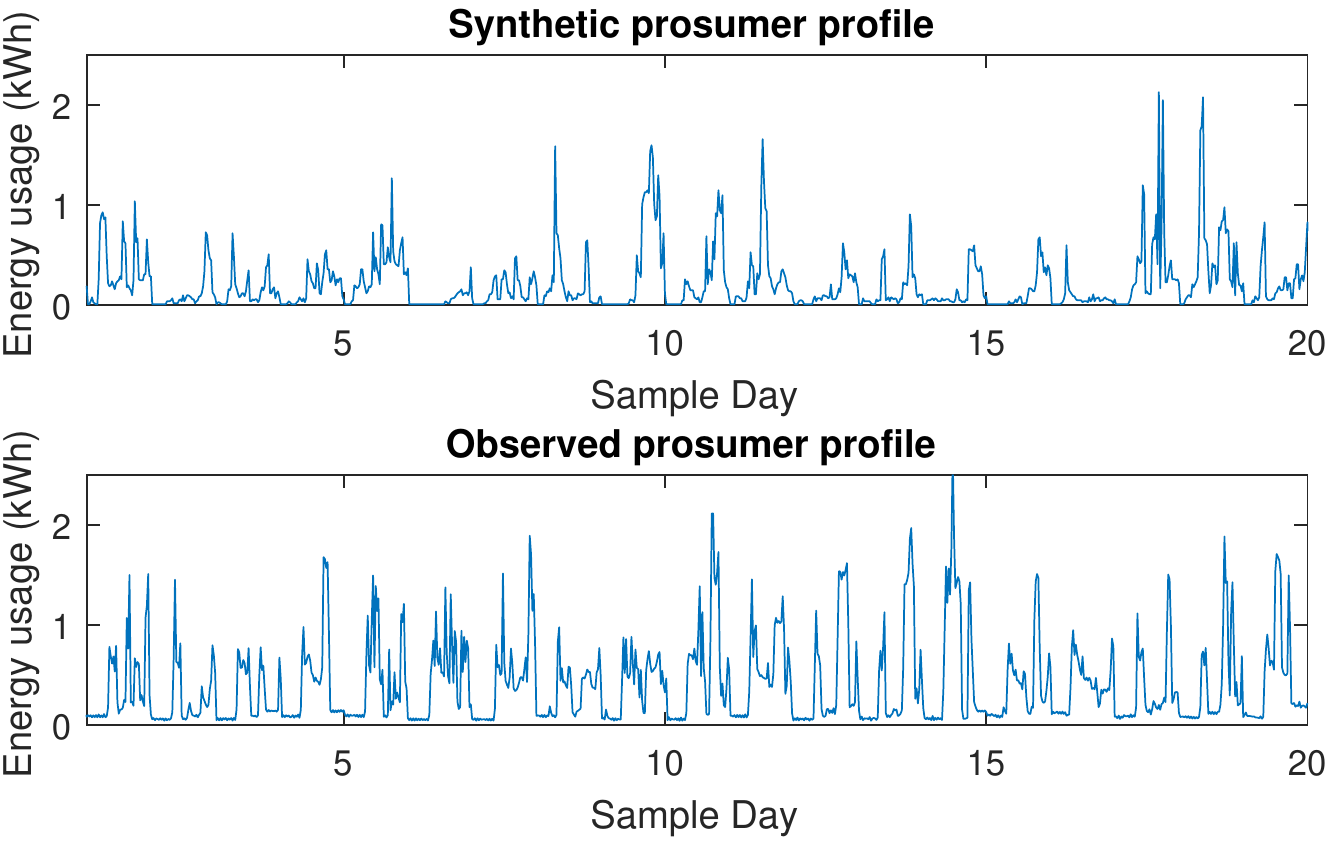}
\caption{Synthetic (top) and observed (bottom) multi-day prosumer profile.} \label{fig:generated_multi}
\vspace{0.1cm}
\end{figure}

Fig. \ref{fig:generated_multi} shows two multi-day prosumer profiles: a synthetic one (top) and a profile observed in the SGSC dataset (bottom).
Some degree of habitual daily behavior is evident in each of these profiles.
To quantify this effect, autocorrelation is calculated for each sample, which measures the correlation between each day and the day preceding it:
\begin{align}
R(\textbf{x}, n) = \frac{1}{(N-n)s^2}\sum_{t=1}^{N-n}(x_t-\bar{x})(x_{t+n}-\bar{x}),
\end{align}
where $n$ is the lag in timesteps, $s^2$ is the sample variance, and $N$ is the total number of samples. The autocorrelation in this case calculates the correlation between any sample $x_{t+n}$ and the corresponding sample $x_t$ at $n$ steps beforehand. Using $n=48$ allows for the correlation between daily behaviors to be quantified. In this particular case, the observed profile had a calculated daily autocorrelation of 0.3066, while the synthetic profile was 0.1993. This shows that in this case, the synthetic profile does emulate habitual daily behavior to a large extent, but does not entirely reproduce the habitual behavior in the observed sample. This is also evident in an average autocorrelation taken over ten synthetic samples and ten observed samples, which yielded 0.1429 and 0.3584 respectively.

\subsection{Generation Model Validation}
\begin{figure}
\centering
\vspace{0.0cm}
\includegraphics[width= \columnwidth]{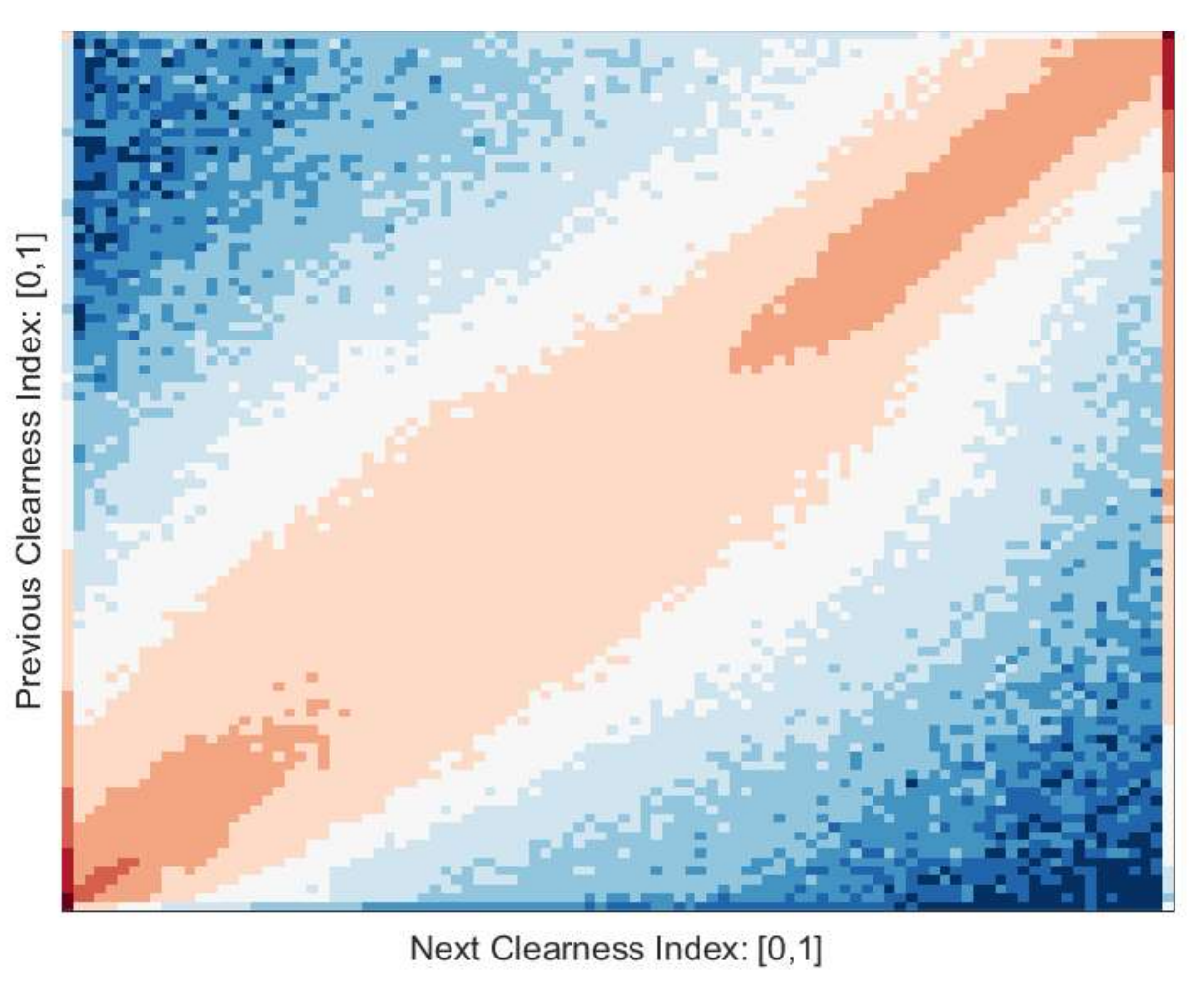}
\caption{Clearness index state transition probability heatmap for 90 observed prosumers.} \label{fig:clearness_index}
\vspace{0.0cm}
\end{figure}

Similarly to the demand model, the generation model is cross-validated using the SGSC data set. Using the methodology outlined in previous work \cite{GV_TP_AUPEC}, a clearness index state transition matrix is generated using data from 90 prosumers over the three-year sample period. This is illustrated in Fig.~\ref{fig:clearness_index}, in which a number of key features are evident.
Firstly, it is apparent that the distribution of clearness index state transitions is bi-modal, with transitions being more likely from low indices to low indices, and high indices to high indices, shown in red. There is also a discontinuity at clearness indices 0 and 1 exactly, at the extreme left and right columns of the figure respectively. Transitions to these states are relatively likely, and transitions from these states are relatively unlikely. This shows that the clearness index has a tendency to transition towards 0 and 1.

\begin{figure}
\centering
\vspace{0.0cm}
\includegraphics[width= \columnwidth]{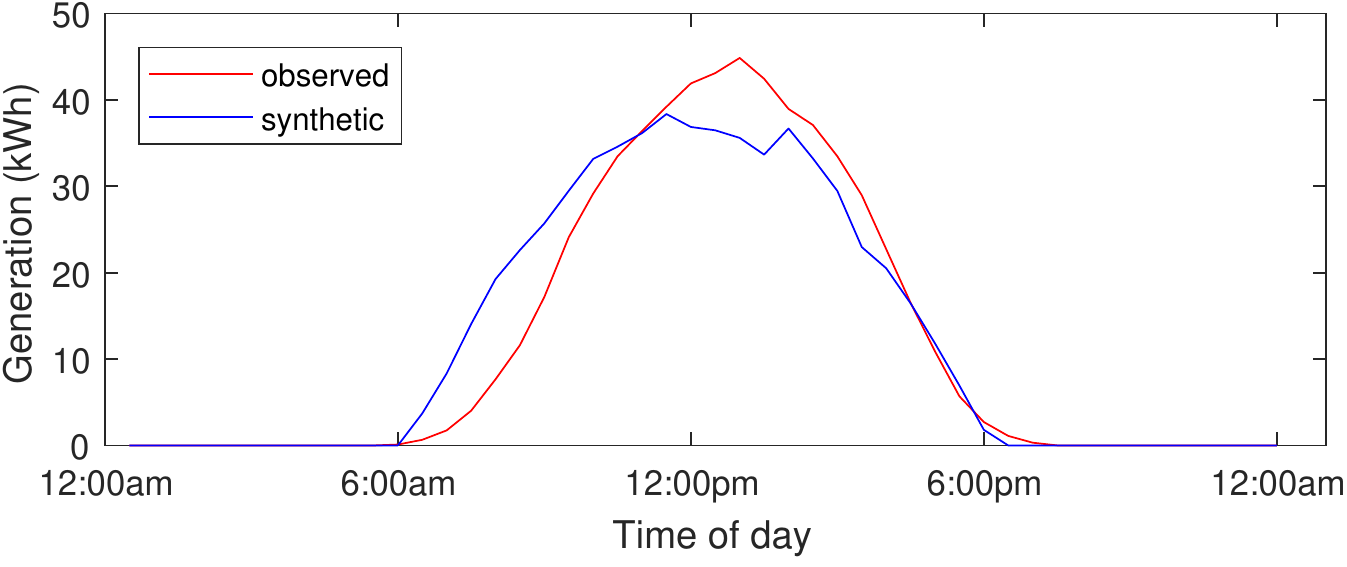}
\caption{100 aggregated synthetic solar profiles (blue), and 100 aggregated observed solar profiles (red).} \label{fig:gen_aggregate}
\vspace{0.0cm}
\end{figure}

Following the population of the clearness index state transition matrix, profiles can be synthesized. In Fig.~\ref{fig:gen_aggregate}, the aggregate of 100 synthetic profiles is shown in red, generated for the approximate location of Sydney city in March. These profiles have been generated similarly to those detailed in \cite{GV_TP_AUPEC}. These are shown along with 100 aggregated observed profiles, which are shown in red.
From Fig. \ref{fig:gen_aggregate}, it is clear that the synthetic solar profiles follow a similar curve to the observed profiles over the course of the day. Given that only 100 profiles were used in this case, the variance is higher than in the demand model cross validation mentioned previously. The maximum absolute error in the estimate occurs at 8:00am, and is \SI{11.6}{kWh}. Maximum absolute error is used in this case as the mean absolute error metric defined for the generation module cross validation is not effective in this case. This is due to the fact that the values approach zero before approximately 6:00am and after approximately 6:00pm, and the percentage error in these cases does not effectively represent the accuracy of the estimation.

\section{Case Studies}
In order to demonstrate some of the potential applications of this model, two specific case studies have been conducted.
The first one makes use of the variation in geography evident in the SGSC data set. The model is used to investigate if there are structural behavioral differences between demand and generation profiles in dense urban, suburban, and rural settings and compares and contrasts these. The second case study investigates the effect on aggregated generation and net demand profiles of an increasing penetration of distributed generation systems.
\subsection{Case Study: Geographical Variation}
\begin{figure}
\centering
\includegraphics[width= \columnwidth]{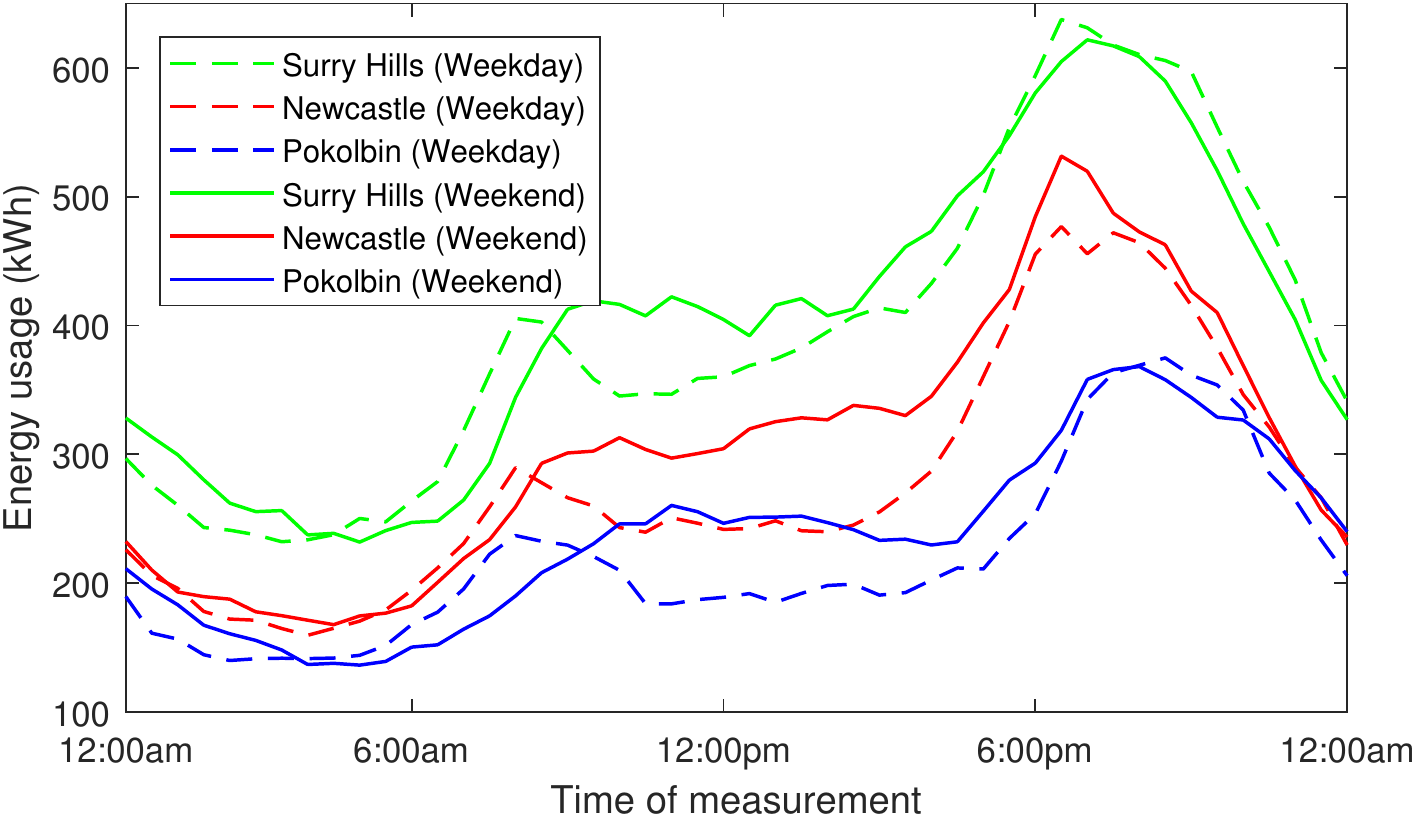}
\caption{1000 aggregated weekday and weekend synthetic demand profiles for Surry Hills (Blue), Newcastle (Red) and Pokolbin (Green).} \label{fig:cs1_p1}
\end{figure}

The first case study illustrates differences in demand behavior between rural, urban and suburban postcodes on both weekdays and weekends. For this study, three postcodes were selected. These are the urban Sydney suburb of Surry Hills, with 276 observed prosumers, Suburban Newcastle, with 832 prosumers, and the rural town of Pokolbin, with 266 prosumers. Using data from these prosumers, demand state transition matrices were compiled for each postcode and 1000 synthetic profiles were generated for each. The aggregated results for weekday and weekend demand are detailed in Fig. \ref{fig:cs1_p1}. It is apparent from this figure that the 1000 simulated residences in Pokolbin (green) have a generally higher demand than those in Newcastle (red) and Surry Hills (blue). It is evident that the morning peak in each profile occurs at approximately the same time, being 7:00am - 8:00am. There is a difference, however, in the afternoon peak, which occurs just after 6:00pm for Pokolbin, with Newcastle peaking just after, and Surry Hills peaking later, at about 8:00pm. In addition to this, the Pokolbin profile appears to have a much bigger morning peak relative to the level of demand over the middle of the day, while the morning peak in Newcastle and Surry Hills is lower.

During the weekend, the demand during the middle of the day is higher. Both Newcastle and Pokolbin have a morning peak at a similar time to weekdays, however Surry Hills has a lowered morning peak, which occurs at around 10:00am-11:00am. For each of the three profiles, demand does not drop significantly after the morning peak, but instead stabilises, or rises throughout the day. In contrast to this, the evening peak and overnight profile does not differ significantly from the weekday profiles.

This case study demonstrates the application of this model to the generation of load profiles in specific areas. The model was therefore able to synthesize 1000 daily profiles for each postcode, using less than 300 prosumers to populate the Pokolbin and Surry Hills models, and approximately 830 prosumers to populate the Newcastle model. This method has applications to the analysis of aggregated loads at the ends of distribution feeders, or even certain zone substations, where only a limited amount of observed data are available.

\subsection{Case Study: Increasing Solar Penetration}
The purpose of this case study is to demonstrate the function of the model to simulate the distribution of solar generation amongst a population, given a small set of observed data, and to investigate the effect on net demand profiles with increasing penetrations of installed solar generation. For this case study, the postcode 2280 is used, this corresponds to the southern suburbs in Newcastle. This postcode is an excellent candidate for this study as it has a large amount of observed demand data available in the SGSC dataset, and in addition, also has 61 prosumers with observed solar data available. This means that of the 225 observed customers with data available, 27.11\% of these have installed solar generation. For this case study, a set of 1000 unobservable prosumers is used.

In the sampling of the aggregate solar load profile, the same clearness index transition profile was used for every prosumer, as they are tightly grouped geographically. This replicates the effect of the suburb experiencing the similar transitions in cloud cover. Following the formulation of an aggregate generation profile, this is subtracted from the demand profile to yield a net demand profile. 

\begin{figure}
\centering
\vspace{0.0cm}
\includegraphics[width= \columnwidth]{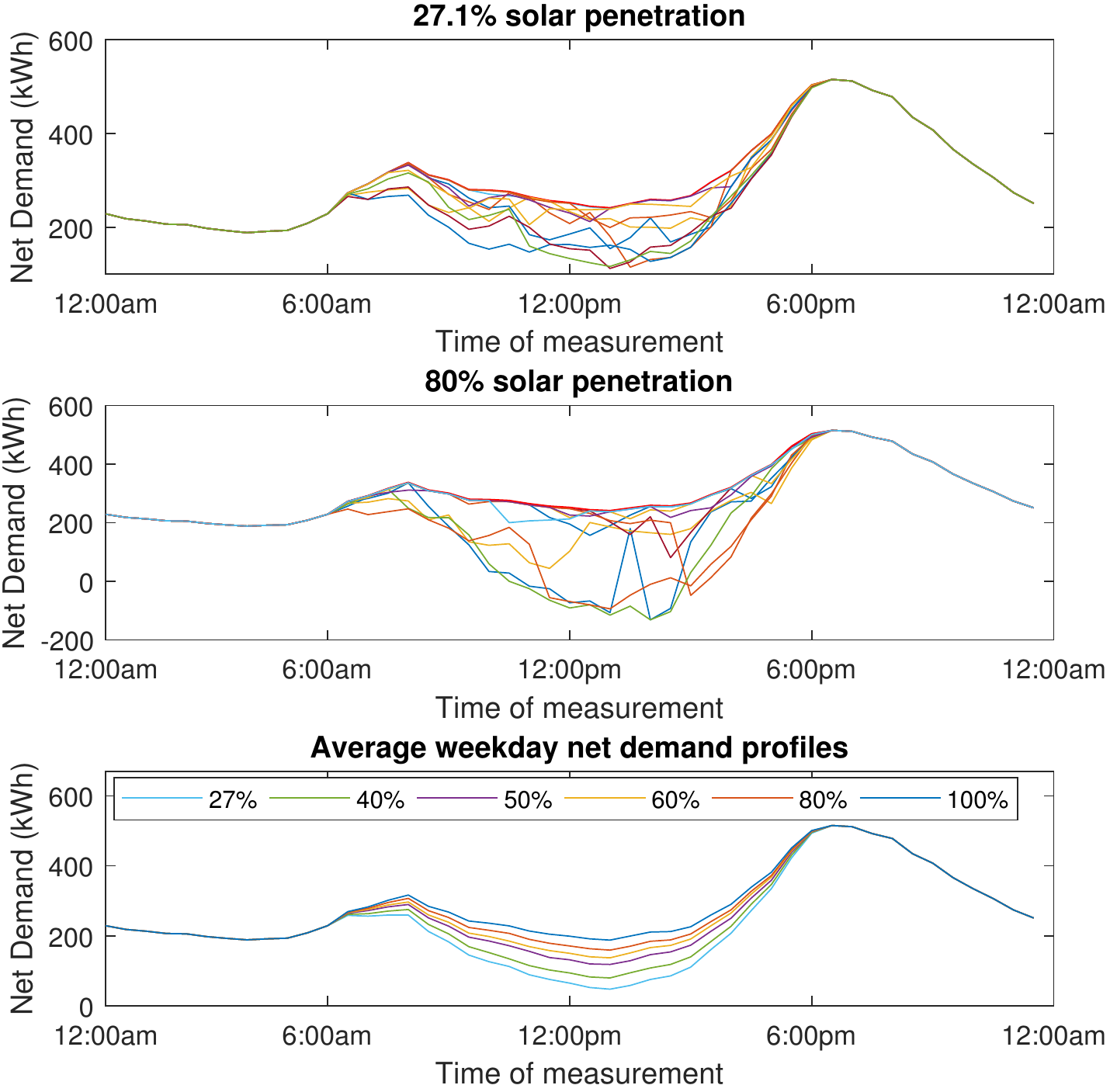}
\caption{The impact of PV penetration on the demand profiles: individual demand profiles for two different PV penetration levels, 27.1\% (top) and 80\% (middle), and the average of 1000 demand profiles for six different PV penetration levels (bottom).} \label{fig:cs2_27}
\vspace{0.0cm}
\end{figure}

Fig. \ref{fig:cs2_27} shows a set of ten synthetic demand profiles for a distributed generation penetration of 27.1\% (top), 80\% (middle) and the average for six different PV penetration levels (bottom). Observe that for the 80\% penetration level, the midday trough is more pronounced, with net demand becoming negative on some days.
One feature which remains common to these charts is that the evening peak is largely unchanged from its gross demand value. It is therefore evident that the level of PV penetration has little impact on the daily peak residential demand, which is the evening peak. In this case study, the solar profiles are generated for early March, and as such it is expected that the evening peak will be similarly unaffected from March to September.

The bottom plot shows the average net demand profile for, respectively, 27\%, 40\%, 50\%, 60\%, 80\% and 100\% PV penetration. Observe that unlike for the individual demand profiles, the average demand profile never goes negative, even for the 100\% PV penetration, which highlights the importance of individual demand profiles.

\section{Conclusion}
In this paper, we have proposed a Bayesian framework for probabilistic generation of synthetic residential demand and generation data using a limited set of observed customer data. The use of the Dirichlet multinomial distribution overcomes a limitation of previous models, which is they do not account for variance in the distribution of characteristics across unobservable customers. We have shown that the methodology performs well even when the number of unobservable customers greatly exceeds the number of observable customers. 
One of the issues with sparse observed data is that some states could be calculated to have a zero probability of being sampled, because they have not been observed in the input data, which we overcome by using Gaussian kernel density estimation. 
Finally, we have proposed a Dirichlet process which allows for the generation of samples that have a tendency to follow recurring trajectories and sequences of states. This sampling process involves reinforcement, where the state sampled is added into the initial matrix of counts. This results in recurring behavior, which is particularly important in synthesizing residential consumption data.
The proposed methodology has been successfully applied to study the impact of network tariffs in low voltage distribution networks \cite{Azuatalam2018} and the impact of PV-battery system impacts on low-voltage distribution networks \cite{Ma2018}, respectively.

\ifCLASSOPTIONcaptionsoff
  \newpage
\fi

\bibliographystyle{IEEEtran}
\bibliography{GenDemModel}

\end{document}